\documentclass{article}

\PassOptionsToPackage{numbers, compress}{natbib} 
\usepackage[preprint]{neurips_2025}

\usepackage[utf8]{inputenc} % allow utf-8 input
\usepackage[T1]{fontenc}    % use 8-bit T1 fonts
\usepackage{hyperref}       % hyperlinks
\usepackage{url}            % simple URL typesetting
\usepackage{booktabs}       % professional-quality tables
\usepackage{amsfonts}       % blackboard math symbols
\usepackage{nicefrac}       % compact symbols for 1/2, etc.
\usepackage{microtype}      % microtypography
\usepackage{xcolor}         % colors

\usepackage{url}
\usepackage{booktabs}
\usepackage{adjustbox}
\usepackage{xspace}

\usepackage{enumitem}
\usepackage{afterpage}

\usepackage{listings}
\usepackage{amsmath}
\usepackage{xspace}
\usepackage{tcolorbox}
\usepackage{listings}
\usepackage{xcolor}
\usepackage{titlesec}
\usepackage{enumitem}
\usepackage{algorithm}
\usepackage{algpseudocode}
\usepackage{adjustbox}
\usepackage{booktabs}
\usepackage{hyperref}
\usepackage[normalem]{ulem}
\usepackage{colortbl }
\usepackage{multirow}
\usepackage{hhline}
\usepackage{pdflscape}
\usepackage{pifont}
\usepackage{latexsym}
\usepackage{stmaryrd}
\usepackage{balance}
\usepackage{caption}
\usepackage{wasysym}
\usepackage[flushmargin,bottom]{footmisc}
\usepackage{comment}
\usepackage{tikz}
\usepackage{float}
\usepackage{amsmath}
\usepackage{amssymb}
\usepackage{dashrule}
\usepackage{verbatim}
\usepackage{authblk}
\usepackage{pifont}
\usepackage{graphicx}
\usepackage{hyperref}
\usepackage{fontawesome5}

\newcommand{\textie}{\textit{i.e.,}\xspace}
\newcommand{\texteg}{\textit{e.g.,}\xspace}

\newcommand{\name}{\textsc{OSS-Bench}\xspace}
\newcommand{\namephp}{\textsc{OSS-Bench$_{php}$}\xspace}
\newcommand{\namesql}{\textsc{OSS-Bench$_{sql}$}\xspace}

\title{OSS-Bench: Benchmark Generator for Coding LLMs}

\author{{\fontsize{11.5pt}{11pt}\selectfont Yuancheng Jiang}}
\author{{\fontsize{11.5pt}{11pt}\selectfont Roland H. C. Yap}}
\author{{\fontsize{11.5pt}{11pt}\selectfont Zhenkai Liang}}
\affil{\textit{School of Computing, National University of Singapore}}
\affil{\small\textsl{\{yuancheng, ryap, liangzk\}@comp.nus.edu.sg}\normalsize}

\begin{document}

\maketitle

\vspace{-10mm}

\begin{center}
  \raisebox{-0.2ex}{%
    \tcbox[
      colback=white, colframe=white, boxrule=0pt, arc=0mm,
      left=1pt, right=1pt
    ]{%
      \includegraphics[height=1em]{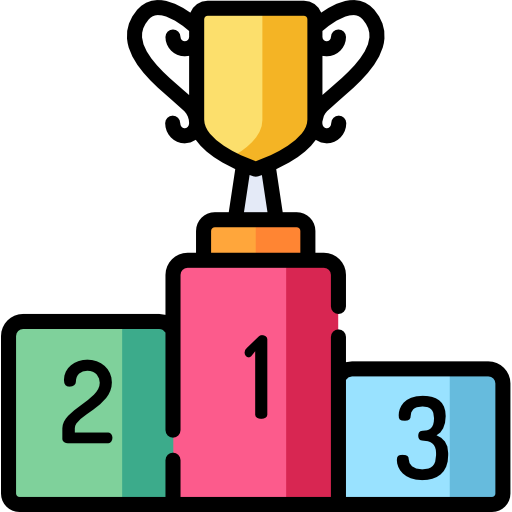}\,%
      \href{https://oss-bench.github.io}{\textbf{Leaderboard}\xspace\xspace\xspace\xspace\xspace}%
    }%
  }%
  \hspace{-0.5em}%
  \raisebox{-0.2ex}{%
    \tcbox[
      colback=white, colframe=white, boxrule=0pt, arc=0mm,
      left=1pt, right=1pt
    ]{%
      \includegraphics[height=1em]{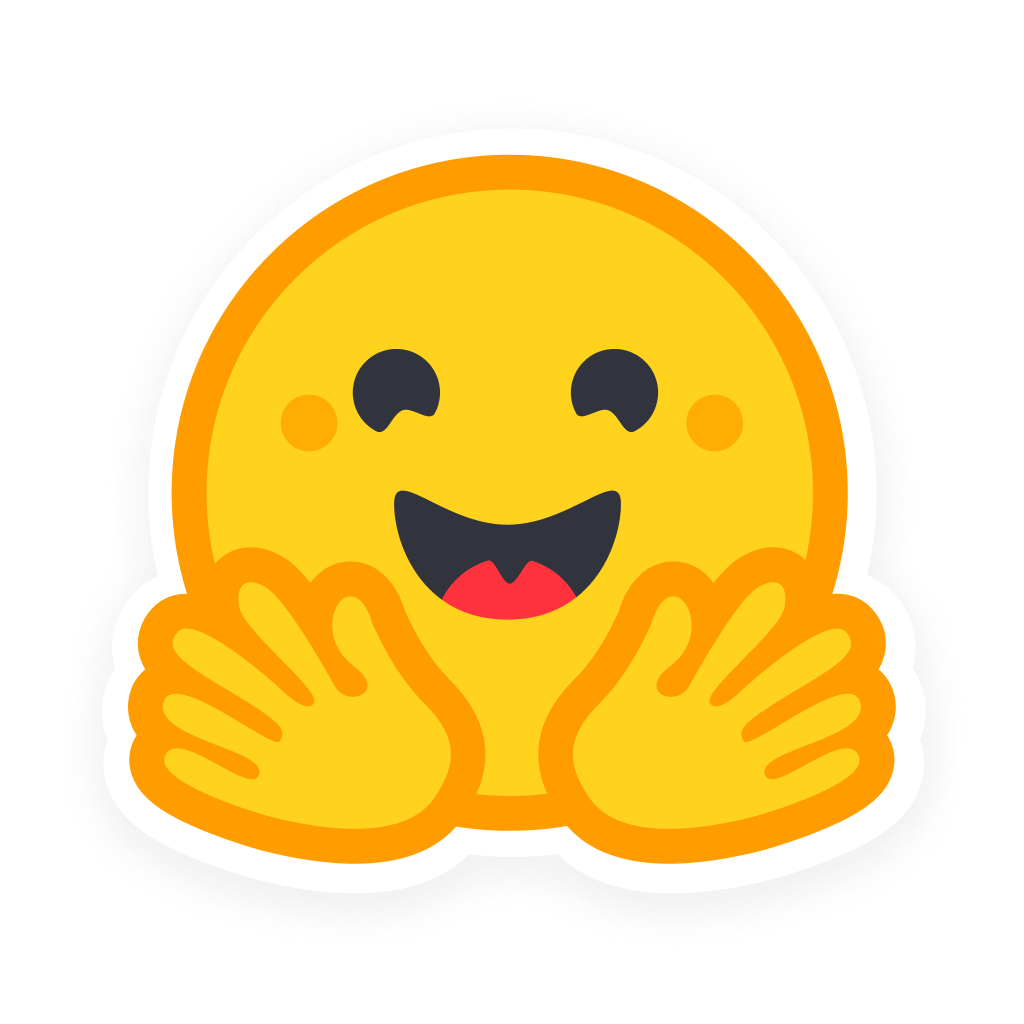}\,%
      \href{https://huggingface.co/datasets/0599jiangyc/oss-bench}{\textbf{Dataset}\xspace\xspace\xspace\xspace\xspace}%
    }%
  }%
  \hspace{-0.5em}%
  \raisebox{-0.3ex}{%
    \tcbox[
      colback=white, colframe=white, boxrule=0pt, arc=0mm,
      left=1pt, right=1pt
    ]{%
      \includegraphics[height=1em]{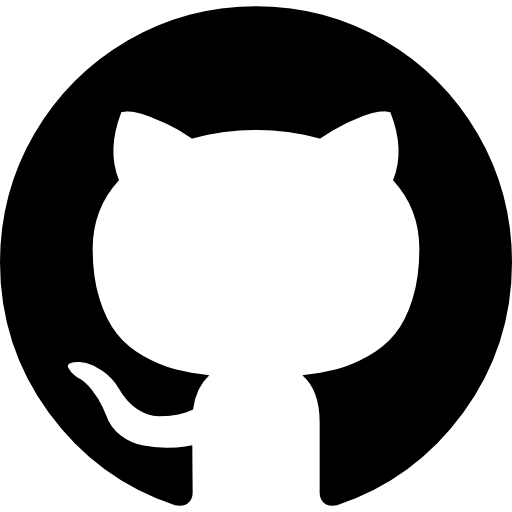}\,%
      \href{https://github.com/oss-bench/oss-bench}{\textbf{Github}}%
    }%
  }%
\end{center}

\begin{abstract}

In light of the rapid adoption of AI coding assistants, LLM-assisted development has become increasingly prevalent, creating an urgent need for robust evaluation of generated code quality. 
Existing benchmarks often require extensive manual effort to create static datasets, rely on indirect or insufficiently challenging tasks, depend on non-scalable ground truth, or neglect critical low-level security evaluations, particularly memory-safety issues. In this work, we introduce \name, a benchmark generator that automatically constructs large-scale, live evaluation tasks from real-world open-source software. \name replaces functions with LLM-generated code and evaluates them using three natural metrics: compilability, functional correctness, and memory safety, leveraging robust signals like compilation failures, test-suite violations, and sanitizer alerts as ground truth. In our evaluation, the benchmark, instantiated as \namephp and \namesql, profiles 17 diverse LLMs, revealing insights such as intra-family behavioral patterns and inconsistencies between model size and performance. Our results demonstrate that \name mitigates overfitting by leveraging the evolving complexity of OSS and highlights LLMs' limited understanding of low-level code security via extended fuzzing experiments. Overall, \name offers a practical and scalable framework for benchmarking the real-world coding capabilities of LLMs.

\end{abstract}

\section{Introduction}
\label{sec:intro}

% \yc{Nine Page before Ref. Appendix allowed. Dataset track - Can be Single/Double Blind} \yc{this is preprint format}

% \ry{ToCheck: 1. Does LLM fix any original bugs found in eval?; 2. Check if LLM already has memorized some PHP codebase, e.g. give code fragment and ask LLM what is the code from - make sure no PHP in names/comments}

Coding with Large Language Model (LLM) assistants such as Copilot~\cite{githubCopilot2021} and Cursor~\cite{cursor2023} is revolutionizing software development by enabling users to refine code snippets through natural language. 
% Rather than relying solely on traditional peer review, LLMs increasingly serve as automated quality inspectors, from assisting with individual daily tasks to reviewing large-scale project development. 
Numerous coding LLMs have emerged in recent years, each claiming strong capabilities across various programming tasks and supporting these claims with competitive or state-of-the-art benchmark results. This rapid growth has created an urgent need for rigorous and robust evaluation of the quality of LLM-generated code.

Numerous benchmarks~\cite{liu2023your, hajipour2024codelmsec, hu2025dynacode, zheng2025humanevo, wang2025codeif, tang2023ml, jimenez2023swe, jain2024livecodebench, zhuo2024bigcodebench, zan2025multiswebench, dilgren2025secrepobench, peng2025cweval, zheng2024beyond, baars2019codearena, quan2025codeelo, feng2024complexcodeeval, yadav2024pythonsaga, zheng2024well, jimenez2023swebench, du2023classeval, peng2024humanevalxl, zheng2023codegeex} have been proposed to assess and quantify LLM coding capabilities, typically focusing on three key aspects: correctness, efficiency, and security. While these benchmarks provide valuable insights into the coding abilities of various LLMs, they commonly suffer from at least one of the following limitations:

% However, we find that they may not properly address the following challenges: (I) employ useful and challenging real world coding tasks; (II) the benchmarks may have manual curation aspects which mean they cannot easily evolve over time and suffer due to lack of automated maintainability; (III) scalability of ground truth determination; and (IV) limited focus on security and code robustness.
% \ry{may rewrite challenges: TODO list. a. useful real-world task which has challenge; b. is there correct ground truth for task; c. benchmark may be outdated if LLM can learn from it (liveness) + minimum maintainability effort; d. not a single benchmark (generator argument)} \yc{a. is the second one; b. is not challenging; they don't have ground-truth problem; they are reliable but require heavy human effort; c. is the first point; d is not a valid argument for benchmark}
%
\begin{itemize}[topsep=0.2mm,parsep=0.2mm,partopsep=0pt,leftmargin=*]
\setlength{\itemsep}{2pt}\setlength{\parsep}{0pt}\setlength{\parskip}{0pt}
    \item[-] \textit{Heavy human effort and static tasks.} Many benchmarks rely on manually crafted test cases, requiring substantial human expertise and effort. These datasets are typically static and lack extensibility, potentially allowing models to overfit by memorizing or hard-coding solutions.

    \item[-] \textit{Indirect or less challenging tasks.} Several benchmarks use indirect prompting (\texteg needle-searching tasks) or simple coding scenarios, such as basic algorithm implementations, which do not sufficiently challenge rapidly evolving LLMs.

    %\item \textbf{Biased or unreliable ground truth.}
    \item[-] \textit{Limited or non-scalable ground truth.} Manually constructed benchmarks are costly, subjective, error-prone, and fail to scale. Automated alternatives—such as using other LLMs to generate ground truth—introduce additional errors and uncertainties into the evaluation.

    \item[-] \textit{Limited focus on code security.} Most existing benchmarks prioritize correctness/efficiency, but not much attention has been paid to security. In particular, low-level code vulnerabilities (\texteg memory-safety issues) remain largely unexplored in current LLM evaluation frameworks.

\end{itemize}

% Finally, we believe that LLM evaluation should be on tasks which are not only challenging but represent useful real-world programming tasks and should be on varied code.
% \ry{Additional features: varying tasks + difficulty}

We advocate for a fully automated approach to evaluating the coding capabilities of LLMs, without human or LLM involvement in test case generation or ground truth annotation, while preserving practically relevant, sufficiently challenging programming tasks that span a diverse range of real-world code. In this work, we introduce \textbf{\name}, an automatic benchmark generator designed to address the challenges outlined above. \name repurposes realistic tasks from open-source software (OSS) projects that meet specific conditions, enabling the benchmarks to be automatically maintained and continuously updated—that is, to function as a live benchmark—based on the natural evolution of the underlying OSS. To reflect real-world development practices, \name prompts LLMs to perform function-level code edits, a common task for developers. To enhance task complexity, it draws from the rich and mature ecosystem of large-scale OSS projects, encompassing sophisticated components such as garbage collectors, compilers, and memory allocators. Finally, \name establishes robust ground truth without manual intervention or LLM-generated oracles, relying instead on three practical and verifiable evaluation metrics:

% We believe LLM evaluation should test on programming tasks that are both genuinely useful in real-world scenarios and sufficiently challenging, tested across diverse code examples. In this work, we introduce \textbf{\name}, an automatic benchmark generator that addresses the challenges above. We leverage on reusing realistic tasks from open-source software (OSS) with some required conditions and allows the generated benchmarks to be automatically maintained and self-evolving, i.e. a live benchmark, relying on the evolution of the underlying OSS. 
% To mirror real-world development, \name prompts LLMs to perform function-level code edits—an everyday practice for developers. 
% To increase task complexity, it leverages the vast, decades-evolved ecosystem of large OSS projects, which includes intricate components such as garbage collectors, compilers and memory allocators. Finally, \name establishes robust ground truth without manual effort or LLM-generated oracles by relying on the following three practical and verifiable evaluation metrics:

\begin{itemize}[topsep=0.2mm, parsep=0.2mm, partopsep=0pt, leftmargin=*]
\setlength{\itemsep}{2pt}
\setlength{\parsep}{0pt}
\setlength{\parskip}{0pt}
    \item \textbf{Compilability.} This measures whether a modified function compiles successfully; failures signal syntax errors (\texteg invalid keywords) or semantic errors (\texteg references to non-existent variables). 
    \item \textbf{Functional Test.} Software testing serves as a proxy for code correctness. Well-maintained OSS projects typically include comprehensive test suites (such as unit tests, integration tests, end-to-end tests and regression tests) to validate the functionality of code changes.  
    \item \textbf{Memory Safety.} Memory safety is evaluated through sanitizer tools (\texteg ASan~\cite{serebryany2012addresssanitizer}) that detect issues such as buffer overflows and use-after-free errors, which can be attacked or exploited. 
    % \item \textbf{Secure Code.} We evaluate how secure is the code by using sanitizers focusing on memory safety\footnote{Memory safety is evaluated through sanitizer tools (\texteg ASan~\cite{serebryany2012addresssanitizer}) that detect issues such as buffer overflows and use-after-free errors, which can be attacked or exploited.} as the most severe vulnerabilities occur with memory errors. %  
\end{itemize}

We instantiate \name to generate two real-world benchmarks—\namephp and \namesql—built on the PHP~\cite{php-src2025} interpreter and the SQLite3~\cite{sqlite3} engine. Both benchmarks prompt LLMs to \textbf{improve} individual functions and then measure the \textbf{degradation} (\textie new compilation failures, test regressions, or sanitizer alerts in the original code) after replacing the original function with the LLM version. \name delivers (1) continuously updatable tasks with no manual curation, (2) complex, realistic challenges grounded in natural OSS signals without relying on other LLMs, and (3) measurements of low-level memory safety.

We assess \name’s effectiveness and fairness on both \namephp and \namesql using 17 widely used LLMs. Our results show that \name not only profiles compilation, testing, and memory-safety performance accurately, but also uncovers insightful observations, such as inconsistencies between model size and performance, and how increased memorization can lead to counterproductive outputs. Furthermore, we demonstrate that \name’s live, continually updated design mitigates overfitting to fixed datasets and scales naturally to large, evolving codebases.

\section{Approach}

\label{sec:approach}

\begin{figure*}[tb]
\setlength{\abovecaptionskip}{5pt}
\setlength{\belowcaptionskip}{0pt}
\setlength{\intextsep}{0pt}
    \centering
    \includegraphics[scale=0.45]{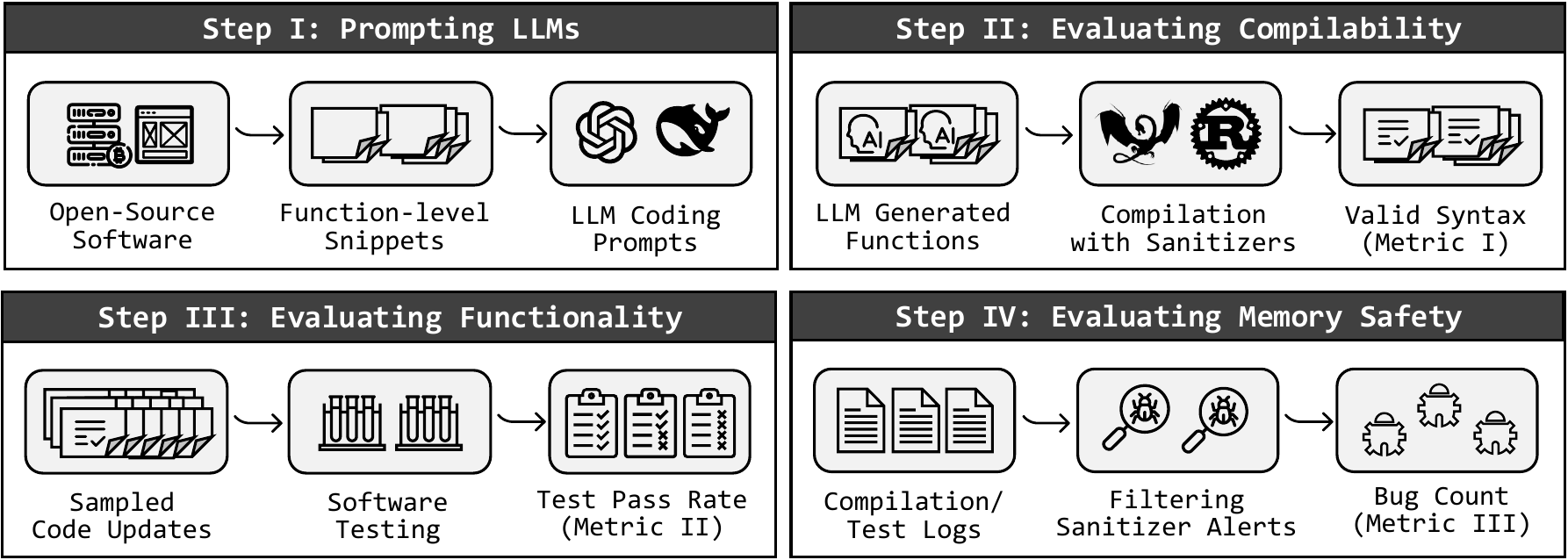}
    \caption{Benchmark Workflow in \name}
    \label{fig:overview}
\end{figure*}

\textbf{Benchmark Workflow.}
Figure~\ref{fig:overview} illustrates \name’s four‐step workflow: (I) select an OSS project, extract function‐level snippets, and prompt LLMs for optimized versions; (II) evaluate the compilability of each LLM‐generated function; (III) sample the successfully compiled functions to form testing datasets and run the OSS test suite; and (IV) collect compilation and test logs and analyse sanitizer alerts for memory‐safety assessment.

% Figure~\ref{fig:overview} presents the proposed benchmark workflow of \name, which consists of four main steps: (i) extracting functions from open-source software and prompting LLMs to generate improved code, (ii) checking the compilability of each improved function from LLMs (iii) sampling and replacing function sets followed by running the software testing, and (iv) performing sanitizer analysis either through quick alert collection or an enhanced mode using fuzzing. 
% \ry{expand on sampling which is code combinations, also harder to learn}

Steps (II), (III), and (IV) correspond to three core metrics for evaluating LLM-generated code quality. A model receives a higher compilation score if more LLM modifications compile successfully; a higher test score if the average test pass rate across sampled iterations is higher; and a higher memory safety score if it triggers fewer sanitizer alerts. In principle, users can apply this workflow to various open-source projects to create diverse benchmarks in different areas. 

\textbf{Open-source Software Criteria.} 
Decades‑old, actively maintained open‑source software provides an ideal foundation for live benchmarks. Consequently, to select projects that offer substantial code complexity, extensive test coverage, and stable release practices, we apply the following criteria:

\begin{itemize}
[topsep=0.2mm,parsep=0.2mm,partopsep=0pt,leftmargin=*]
\setlength{\itemsep}{2pt}
\setlength{\parsep}{0pt}
\setlength{\parskip}{0pt}
    \item[-] \textit{Large codebase.} A large codebase increases the inherent complexity of evaluation tasks on realistic development scenarios. Although the LLM edits a small code snippet, that snippet must integrate seamlessly into the broader, complex codebase. 
    \item[-] \textit{Comprehensive test suite.} The selected OSS must include an actively maintained, well-designed test suite with broad coverage (\texteg 60\% code coverage, the acceptable coverage considered by Google~\cite{google-coverage}). Ideally, the suite combines unit tests and end-to-end tests to validate both individual functions and their interactions across the codebase.
    \item[-] \textit{Compiled languages.} Compiled programming languages (\texteg C, C++, Rust) that require a build step are inherently more complex than scripting languages (\texteg Python, JavaScript). Evaluating on compiled languages (1) increases coding task difficulty and (2) provides natural, robust ground truth via compilability and sanitizer checks.  
\end{itemize}

Given these criteria, two major OSS categories commonly qualify: programming language implementations and database management systems. Programming language implementations (\texteg CPython, Clang, GCC, PHP) have complex codebases with parsers, compilers, garbage collectors, and runtime environments—ideal for testing LLMs on challenging low-level tasks. Database management systems (\texteg MySQL, SQLite, DuckDB) serve as core infrastructure with intricate query optimizers, transaction engines, and storage layers, and their extensive test suites provide rich ground truth for correctness and safety. 

Beyond these, OSS-Bench can be flexibly extended to other complex systems—such as web servers and networking stacks (\texteg Nginx, Apache HTTPD), container/orchestration platforms (\texteg Docker, Kubernetes), and ML frameworks (\texteg TensorFlow C-API, PyTorch’s C++ frontend). These mature, actively maintained projects offer large codebases and comprehensive test suites, making them excellent live datasets for realistic LLM benchmarking.

\textbf{Function Extraction.} \name employs \texttt{libclang}~\cite{libclang}, part of the LLVM toolchain, to reliably extract all function definitions—including comments—from the codebase. Extracted functions are stored in a database and filtered by token count to exclude those that are excessively short or long. In typical large OSS projects, this process yields thousands of functions; by default, we retain functions with 10–256 tokens, filtering out approximately 10\% of the corpus.

\begin{figure*}[tb]
\setlength{\abovecaptionskip}{5pt}
\setlength{\belowcaptionskip}{0pt}
\setlength{\intextsep}{0pt}
    \centering
    \includegraphics[scale=0.42]{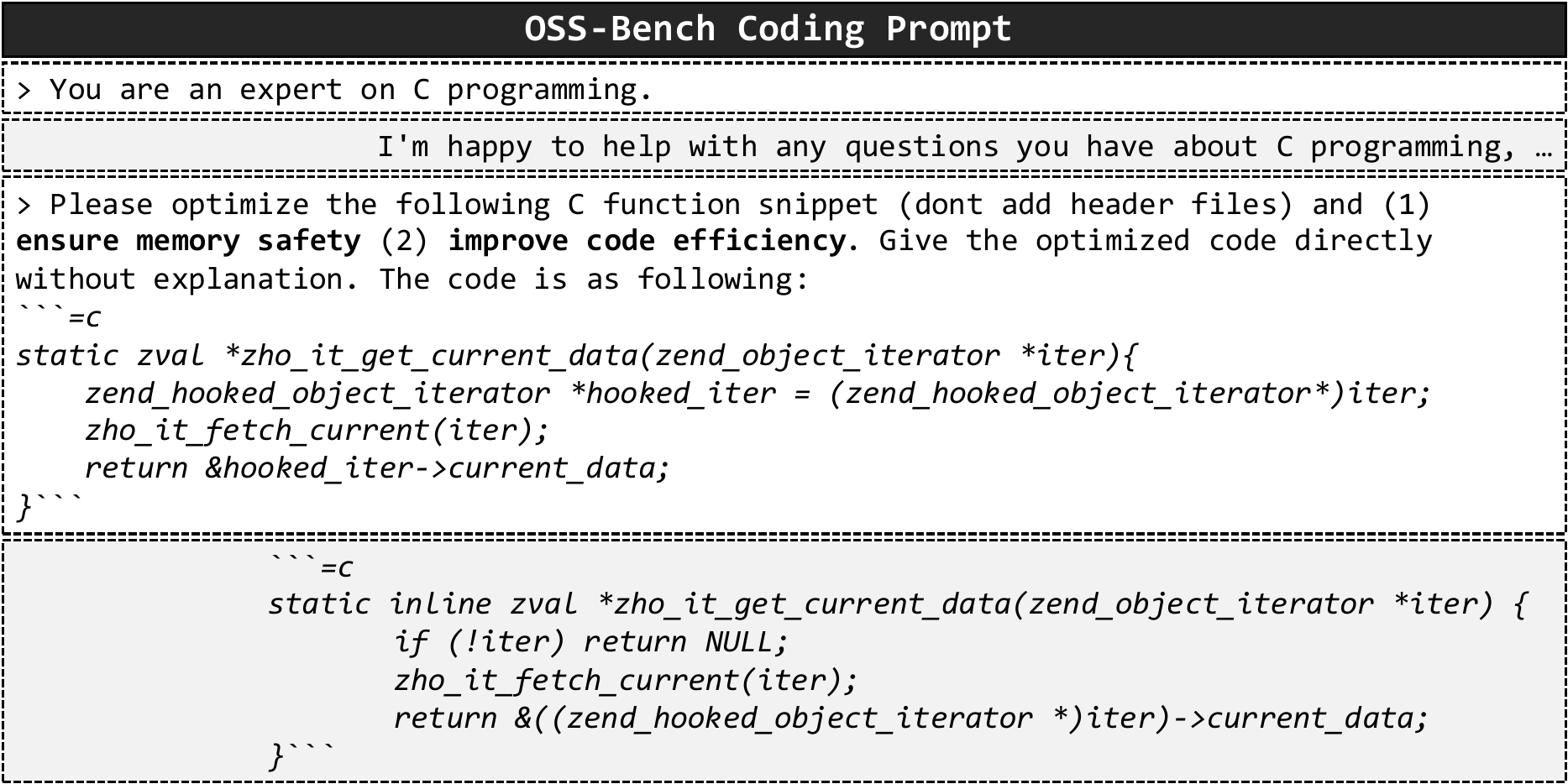}
    \caption{\name uses daily code prompts to simulate common LLM-assisted coding workflows.}
    \vspace{-1mm}
    \label{fig:prompt}
\end{figure*}

\textbf{\name Coding Prompt.} Our prompt emulates common AI-assisted coding workflows: users provide an existing code snippet and request for improvement. As illustrated in Figure~\ref{fig:prompt}, it asks the LLM to improve a given code snippet at the function level with two primary goals: \textit{(1) ensure memory safety} and \textit{(2) enhance code efficiency}. The first objective guarantees a fair evaluation of Metric III (memory safety) by explicitly highlighting the attention to code safety. The second objective drives substantive logic improvements, preventing trivial or minimal edits. \name’s prompt framework is also extensible—supporting other tasks like full code synthesis or automated bug repair.  

\textbf{Metric I --- Compilability.} Compilability is one natural and common task in compiled languages. Incompilability indicates syntax (\texteg incorrect grammar) or semantic (\texteg use of undefined variables) errors. \name first compiles the original open-source software to ensure it builds without errors. 
Then, \name selects each from extracted OSS functions, and replaces the original function with its improved version from language models. After code replacement, \name re-compiles the entire project and check its compilability. 
The compilation result can be success, failure and unrelated (\textie changing such functions do not affect the compilation). \name records each function's compilation result, and the overall compilation pass rate (unrelated functions are excluded) is used as the final score. 

\textbf{Metric II — Functional Testing.} Open-source projects rely on extensive test suites (unit, integration, end-to-end, and regression tests) to validate code updates. To measure this metric, \name first builds the original OSS and runs its official tests to record a baseline. Next, from the set of functions that compiled successfully in Metric I, \name performs 1,000 test iterations: each iteration randomly selects a subset of the compilable functions (\texteg we use $\sim$1\% of the total function number, that is, 100 functions for \namephp and 73 functions for \namesql per iteration) and replaces the originals to simulate an LLM-driven update. The test suite is then executed, and the pass rate is recorded for each iteration. Finally, we compute the average test-pass rate over all iterations and quantify the degradation relative to the baseline.

\textbf{Metric III --- Memory Safety.} Memory safety is a fundamental security concern in compiled languages. Bugs such as buffer overflows and double-free errors are known harmful and can be maliciously exploited. \name aggregates the sanitizer alerts from compilation and test logs during previous metrics, and reports the number of unique memory safety bugs. A lower final count of sanitizer alerts results in a higher Challenge III score. 
Inspired by widely deployed OSS fuzzing frameworks such as OSS-Fuzz~\cite{serebryany2017oss}, \name extends this metric with an in-depth fuzzing evaluation detailed in Section~\ref{sec:extended_fuzz}.
We highlight that the evaluation shows Metric II using test suites is insufficient, as many bugs are observed after the test suite is passed with Metric III.

\textbf{Dissimilarity.} Conservative models may game our benchmark by leaving the original function largely unchanged, thereby inflating their scores. To prevent this, we introduce a \textit{Delta} metric that quantifies the average number of differing lines per function—computed using \texttt{difflib}~\cite{python_difflib2025}—and serves as a configurable impact factor in the final score. Models whose edits fall below a minimum dissimilarity threshold are excluded from evaluation, and higher dissimilarity can enjoy a bonus in the overall score. This ensures that LLMs perform substantive optimizations, in line with our prompt’s emphasis on efficiency improvements, rather than relying on conservative or no changes.

\textbf{Scoring.} Each model will first be evaluated through metric scores with raw score ranging from 0 to 100 as follows: (i) Raw compilability score ($s_1$): the proportion of valid functions that compile successfully after LLM optimization. Valid functions are those whose modifications could affect the build (\textie excluding cases where changes do not alter compilation behavior). We obtain $s_1$ by multiplying this proportion by 100, yielding a raw score between 0 and 100. (ii) Raw test score ($s_2$): the average test pass rate of 1,000 iterations of functional testing. We obtain $s_2$ by multiplying this pass rate by 100, yielding a raw score between 0 and 100. (iii) Raw sanitizer score ($s_3$): begins at 100, with each detected sanitizer alert reducing the score by a certain rate (down to a minimum of 0), so that fewer alerts correspond to a higher score.

\name does not simply sum the three raw scores, as later metrics depend on earlier ones (\textie functionality is evaluated only for compilable code, and memory safety is assessed only for functional code). We incorporate this dependency in the form of \emph{chained scores} (\(c_1, c_2, c_3\)) as:

\vspace{-3mm}
\[
c_1 = s_1 \quad\big|\quad
c_2 = s_2 \times \frac{c_1}{100} \quad\big|\quad
c_3 = s_3 \times \frac{c_2}{100}
\]
\vspace{-3mm}

where \(s_1\), \(s_2\), and \(s_3\) are the raw scores for compilation, testing, and sanitization, respectively. For flexible scoring, we introduce weights \(w_1, w_2, w_3 \in [0,1]\), which default to equal values. To discourage trivial edits, we add a \emph{dissimilarity} bonus \(d\), defined as the average number of changed lines per function, with a default weight \(w_d = 0.1\). The final score \(s\) is:

\vspace{-3mm}
\[
s = w_1\,c_1 + w_2\,c_2 + w_3\,c_3 + w_d\,d.
\]
\vspace{-3mm}

\textbf{Automation and Scalability.} OSS-Bench is highly automated: no manual test-case design or ground-truth annotation is required. The only human effort needed is to write adapters (\texteg compilation scripts, log parsers) for each open-source software. Scalability is achieved through configurable components—project selection, prompt templates, sampling rates, metric weights, and even the addition of new metrics (\texteg static analysis warnings or performance profiles). This flexibility allows \name to scale from single repositories to entire ecosystems, support multiple programming languages, and adapt dynamically as repositories evolve over time.

\section{Evaluation}

\label{sec:evaluation}

\subsection{Evaluation Setup}

\textbf{Selected models}. We select 17 models in total as evaluation candidates, including both closed- and open-source variants, across base and instruction-tuned configurations, spanning various parameter sizes and evaluated under the optimal quantization setting (fp16). These models are \small{\textit{GPT-O1}~\cite{gpt-o12025}, \textit{GPT-O3-Mini}~\cite{o3-mini2025}, \textit{Claude-3.7-Sonnect}~\cite{claude-3-7-sonnet2025}, \textit{Claude-3.5-Haiku}~\cite{claude-3-5-haiku2025}, \textit{Gemini-2.5-Flash}~\cite{gemini-2.5-flash2025}, \textit{Llama3.3-70B-Instruct}~\cite{llama3.3-70b-instruct-fp16}, \textit{CodeLlama-70B-Instruct}~\cite{codellama-70b-instruct-fp16}, \textit{Qwen2.5-Coder-32B-Instruct}~\cite{qwen2.5-coder-32b-instruct-fp16}, \textit{Qwen3-A3B-30B-Instruct}~\cite{qwen3-30b-a3b-fp16}, \textit{Qwen3-8B-Instruct}~\cite{qwen3-8b-fp16}, \textit{Gemma3-27B-Instruct}~\cite{gemma3-27b-it-fp16}, \textit{Qwen2.5-Coder-14B-Instruct}~\cite{qwen2.5-coder-14b-instruct-fp16}, \textit{DeepSeek-V2-Coder-16B-Instruct}~\cite{deepseek-coder-v2-16b-lite-instruct-fp16}, \textit{Starcoder2-15B-Instruct}~\cite{starcoder2-15b-instruct-v0.1-fp16}, \textit{Phi4-14B}~\cite{phi4-14b-fp16}, \textit{Mistral-7B-Instruct}~\cite{mistral-7b-instruct-fp16}, \textit{CodeGemma-7B-Instruct}~\cite{codegemma-7b-instruct-fp16}}, sourced from leading organizations such as OpenAI~\cite{openai2025}, Anthropic~\cite{anthropic2025}, Google~\cite{google2025}, DeepSeek~\cite{deepseek2023}, Alibaba~\cite{alibaba2025}, Meta~\cite{meta2025}, Microsoft~\cite{microsoft2025}, BigCode~\cite{bigcode2025} and Mistral~\cite{mistral2025}.

\textbf{Setup}. We run open-source models using the Ollama~\cite{ollama2025} platform. To ensure reproducibility, we set the random seed to $0$ for all models, except for the Claude series, which does not support custom seed settings. \name provides results on 10,534 functions extracted from the PHP interpreter and 7,321 functions from the SQLite3 database engine. \namephp is tested in commit \small{\textit{3786cff1f3f3d755f346ade78979976fee92bb48}}, \namesql is tested in commit \small{\textit{942c9587698715734715242737dba07ef296b0ef}}. We use \emph{pass@k}~\cite{kulal2019spoc, chen2021evaluating} with \(k=1\) by default to evaluate model performance across the three metrics, given the large number of tasks. Due to limited time and computing resources, we use a subset of models in \namesql.

\textbf{Experimental Infrastructure.} Experiments were conducted across multiple hardware configurations: CPUs (AMD EPYC 9184X, Intel Core Ultra 9 285K, AMD Ryzen 9 9950X), GPUs (NVIDIA A40, H100), and system memory ranging from 32 GB to 512 GB. All experiments were performed on Ubuntu 22.04 and Ubuntu 24.04.

% For each open-source software project, we include baseline references from the original implementations (PHP and SQLite3), which achieve the highest scores among all models. This is expected, as these projects are actively

\subsection{Effectiveness of \name}

\definecolor{gold}{RGB}{255,215,0}
\definecolor{silver}{RGB}{192,192,192}
\definecolor{bronze}{RGB}{205,127,50}

\begin{table}[!t]
\centering
\caption{
Overall results (degradation ranking) of \namephp and \namesql compared to the baseline open-source projects. Note: * indicates the \small{\textit{instruct}} models; thinking is not enabled in all open-source models; all open-source models are under \small{\textit{fp16}} quantization.}
\vspace{1mm}
\scalebox{0.84}{
\begin{tabular}{lcccccc}\toprule
Models &Param. Size & Compilability & \xspace Func. Test\xspace & Mem. Safety & Delta (10\%) & \xspace Score\xspace \\\midrule
\multicolumn{7}{c}{\textbf{\namephp --- The PHP Interpreter}} \\
\midrule
\rowcolor{gray!10}
PHP (baseline)            & N/A & 100   & 99.4  & 100  & 0    & 99.6 \\
\textcolor{gold}{\faMedal} Claude-3.7-Sonnet       & N/A & \small{\textcolor{silver}{\faMedal}}\textbf{92.97} & \small{\textcolor{silver}{\faMedal}}\textbf{89.61} & \small{\textcolor{bronze}{\faMedal}}\textbf{84.13} & 3.18 & 85.3 \\
\textcolor{silver}{\faMedal} GPT-O1                    & N/A & \small{\textcolor{gold}{\faMedal}}\textbf{93.45}   & \small{\textcolor{bronze}{\faMedal}}\textbf{86.81} & \small{\textcolor{gold}{\faMedal}}\textbf{89.65}   & 2.52 & 85.0 \\
\textcolor{bronze}{\faMedal} GPT-O3-Mini               & N/A & 92.72                                           & \small{\textcolor{gold}{\faMedal}}\textbf{89.80} & \small{\textcolor{silver}{\faMedal}}\textbf{84.48} & 2.43 & 84.5 \\
Qwen3-A3B           & 30B & \small{\textcolor{bronze}{\faMedal}}\textbf{92.90} & 82.71                                           & 68.95                                       & 1.11 & 75.3 \\
Gemini-2.5-Flash      & N/A & 88.27                                           & 76.12                                           & 59.29                                       & 1.66 & 66.8 \\
Qwen3                     & 8B  & 88.36                                           & 74.93                                           & 55.84                                       & 1.52 & 65.4 \\
Claude-3.5-Haiku      & N/A & 84.47                                           & 72.77                                           & 53.77                                       & 3.16 & 62.8 \\
Qwen2.5-Coder* & 32B & 83.50                                           & 68.79                                           & 38.25                                       & 2.50 & 56.8 \\
Qwen2.5-Coder* & 14B & 79.08                                           & 64.41                                           & 34.11                                       & 2.10 & 51.2 \\
Llama3.3*   & 70B & 70.89                                           & 62.17                                           & 47.56                                       & 3.40 & 48.7 \\
Gemma3*             & 27B & 71.32                                           & 63.28                                           & 30.66                                       & 1.96 & 45.4 \\
StarCoder2* & 15B & 73.35                                           & 56.81                                           & 12.37                                       & 2.54 & 42.6 \\
Phi4                     & 14B & 66.59                                           & 58.53                                           & 25.14                                       & 3.12 & 41.6 \\
DeepSeek-Coder-V2* & 16B & 67.06                                           & 44.17                                           & 23.76                                       & 2.09 & 36.7 \\
CodeGemma*   & 7B  & 65.48                                           & 41.76                                           & 12.37                                       & 1.55 & 33.6 \\
CodeLlama* & 70B & 53.52                                           & 34.53                                           & 18.93                                       & 2.67 & 27.8 \\
Mistral*       & 7B  & 41.26                                           & 28.26                                           & 0.01                                        & 2.87 & 20.5 \\ 
\midrule
\multicolumn{7}{c}{\textbf{\namesql --- SQLite3 Database Engine}} \\
\midrule
\rowcolor{gray!15}
SQLite (baseline)                 & N/A & 100   & 100   & 100   & 0    & 100  \\
\textcolor{gold}{\faMedal} GPT-O3-Mini            & N/A & \small{\textcolor{gold}{\faMedal}}\textbf{96.68} & \small{\textcolor{gold}{\faMedal}}\textbf{91.81} & \small{\textcolor{gold}{\faMedal}}\textbf{72.16} & 2.38 & 85.5 \\
\textcolor{silver}{\faMedal} Qwen3-A3B & 30B & \small{\textcolor{silver}{\faMedal}}\textbf{96.34} & \small{\textcolor{silver}{\faMedal}}\textbf{72.03} & \small{\textcolor{silver}{\faMedal}}\textbf{55.92} & 1.50 & 69.7 \\
\textcolor{bronze}{\faMedal} Qwen3               & 8B  & \small{\textcolor{bronze}{\faMedal}}\textbf{94.80} & \small{\textcolor{bronze}{\faMedal}}\textbf{69.83} & \small{\textcolor{bronze}{\faMedal}}\textbf{49.83} & 1.59 & 66.3 \\
Claude-3.5-Haiku       & N/A & 92.59 & 60.65 & 37.50 & 2.61 & 59.2 \\
Qwen2.5-Coder*     & 32B & 92.47 & 55.04 & 33.59 & 1.85 & 55.3 \\
Qwen2.5-Coder*     & 14B & 90.31 & 49.58 & 35.04 & 1.77 & 52.0 \\
Phi4               & 14B & 78.71 & 39.31 & 35.91 & 2.91 & 43.2 \\
Deepseek-Coder-V2* & 16B & 82.74 & 45.46 &  7.20 & 1.98 & 43.0 \\
StarCoder2*        & 15B & 81.61 & 41.33 &  0.01 & 1.91 & 40.4 \\
\bottomrule
\end{tabular}
}
\label{tab:results}
\end{table}

\begin{figure*}[t]
\setlength{\abovecaptionskip}{5pt}
\setlength{\belowcaptionskip}{0pt}
\setlength{\intextsep}{0pt}
    \centering
    \includegraphics[scale=0.7]{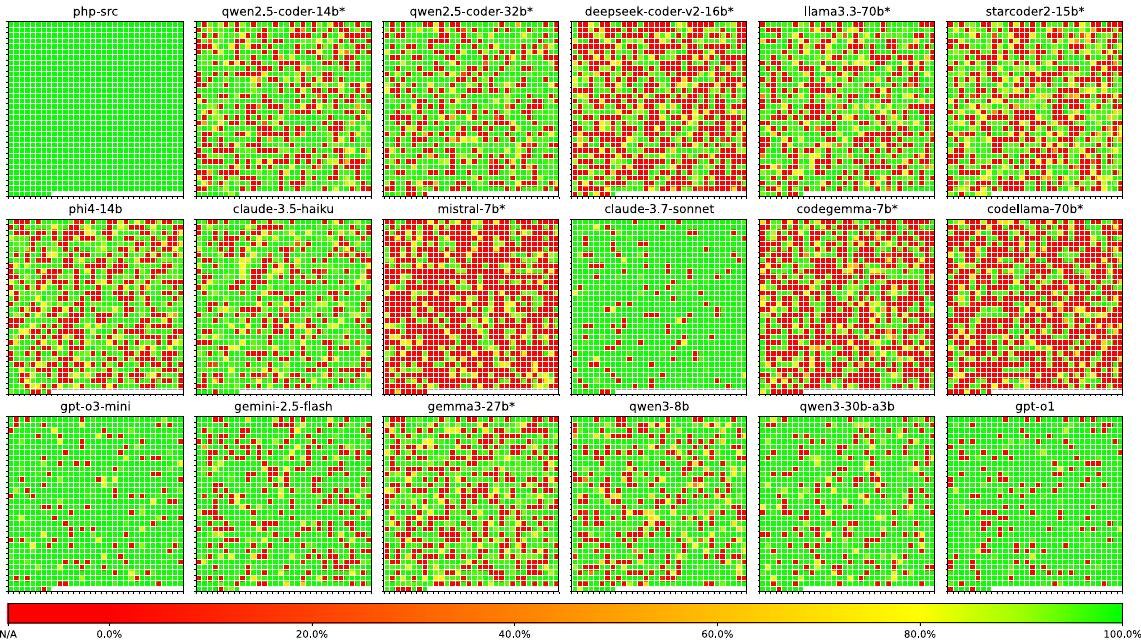}
    \caption{Visualized Test Pass Rates of 1,000 Test Iterations in \namephp}
    \label{fig:test}
\end{figure*}

% We evaluate \name to show how effectively does \name assess the coding capabilities of popular large language models via comparability, test, and memory safety metrics.

% \name profiles LLMs through three metrics (1) syntax and semantics correctness via compilability (2) functionality correctness via functional tests and (3) memory safety via sanitizer checks. We first analyze the overall performance of \namephp shown in Table~\ref{tab:results}. 

We show our benchmarks, \namephp and \namesql, can effectively profile LLM coding capabilities and uncover insightful observations through evaluating three metrics proposed in \name.

\textbf{Overall Results.}\footnote{More detailed results are presented interactively at \url{https://oss-bench.github.io/}.} Table~\ref{tab:results} summarizes the evaluation of \namephp and \namesql, reporting raw scores \(s_1\), \(s_2\), and \(s_3\) for compilability, testing, and sanitization, respectively, along with the dissimilarity bonus \(d\) and the final chained score \(s\). The penalty for each unique sanitizer alert is configurable; by default, we deduct 0.69 points per alert in \namephp and 0.19 points in \namesql, while memory-leak alerts incur only half the standard deduction. Scores are rounded to two decimal places. The original OSS projects (PHP and SQLite3) serve as baselines: all raw scores are 100 after we evaluate the same metrics, except PHP’s test suite, which yields 99.4 due to some expected failing tests. Models that incur less degradation relative to these baselines achieve higher final scores, indicating stronger real-world coding capabilities.

Examining \namephp reveals a clear performance hierarchy. At the top, closed-source, instruction-tuned models consistently achieve the highest compilation and testing scores while generating the fewest sanitizer alerts. A notable outlier is GPT-O3-Mini, which—despite its smaller size—matches or exceeds the performance of larger mid-tier models across all three metrics. Moving down the ranks, mid-tier models begin to show significant degradation in memory safety, even when their compilation and test-pass rates remain respectable. Finally, smaller open-source models exhibit the steepest declines, particularly in triggering memory-safety issues, highlighting that neither parameter count nor general coding ability alone guarantees robust security in complex, real-world codebases.

\textbf{Qwen Family Comparison.} Analyzing the Qwen family on \namephp and \namesql reveals clear performance gains across generations. The third-generation Qwen3 models—especially Qwen3-A3B-30B—consistently outperform their Qwen2.5 predecessors in all three metrics (compilability, functional testing, and memory safety). For example, Qwen3-A3B-30B achieves higher compilation success and test-pass rates than Qwen2.5-Coder-32B, while also triggering fewer sanitizer alerts. Remarkably, even the smallest Qwen3-8B model surpasses the much larger Qwen2.5-Coder-32B across every metric, demonstrating that architectural improvements outweigh mere parameter count. Notably, however, the Qwen3 series tends to make more conservative edits—reflected in lower dissimilarity scores—indicating that these models achieve stronger performance through targeted code modifications rather than extensive rewrites.  

\textbf{Model Size.} Model size does not reliably predict performance. For instance, on \namephp the 70B-parameter CodeLlama scores lower across all three metrics than several lighter models (7B–32B). Likewise, in \namesql the 8B-parameter Qwen3-8B outperforms most other open-source models, highlighting that architecture and training strategy matter more than sheer parameter count in real-world coding tasks.
 
\textbf{Test Result Analysis.} Figure~\ref{fig:test} visualizes the results of 1,000 functional test iterations for \namephp, covering the PHP baseline and 17 LLMs. Each dot represents the average test pass rate for a single iteration, color-coded from red (0\%) to green (100\%). Deep red indicates N/A, corresponding to cases where the test suite fails to start after incorporating LLM-generated code. The visualization clearly differentiates model performance: top performers—GPT‑O1, GPT‑O3‑Mini, and Claude‑3.7‑Sonnet—exhibit predominantly green dots and few failures, whereas models such as Mistral‑7B‑Instruct, CodeGemma‑7B‑Instruct, and CodeLlama‑70B‑Instruct display numerous red and yellow dots, reflecting frequent test failures or low pass rates. 

% Figure~\ref{fig:compile} and Figure~\ref{fig:compile4sqlite} visualize the compilation outcomes in \namephp and \namesql, with each dot indicating a compilation error introduced by an LLM’s code edit. The baseline (\texttt{php-src}, the original PHP interpreter) exhibits zero compilation errors. From the dot density, we observe that closed-source LLMs substantially outperform lightweight and mid-sized open-source models, as they generate far fewer compilation errors.

% \begin{figure*}[t]
% \setlength{\abovecaptionskip}{5pt}
% \setlength{\belowcaptionskip}{0pt}
% \setlength{\intextsep}{0pt}
%     \centering
%     \includegraphics[scale=0.4]{figures/compile.pdf}
%     \caption{Compilation}
%     \label{fig:compile}
% \end{figure*}

\begin{figure*}[b]
\setlength{\abovecaptionskip}{5pt}
\setlength{\belowcaptionskip}{0pt}
\setlength{\intextsep}{0pt}
    \centering
    \includegraphics[scale=0.41]{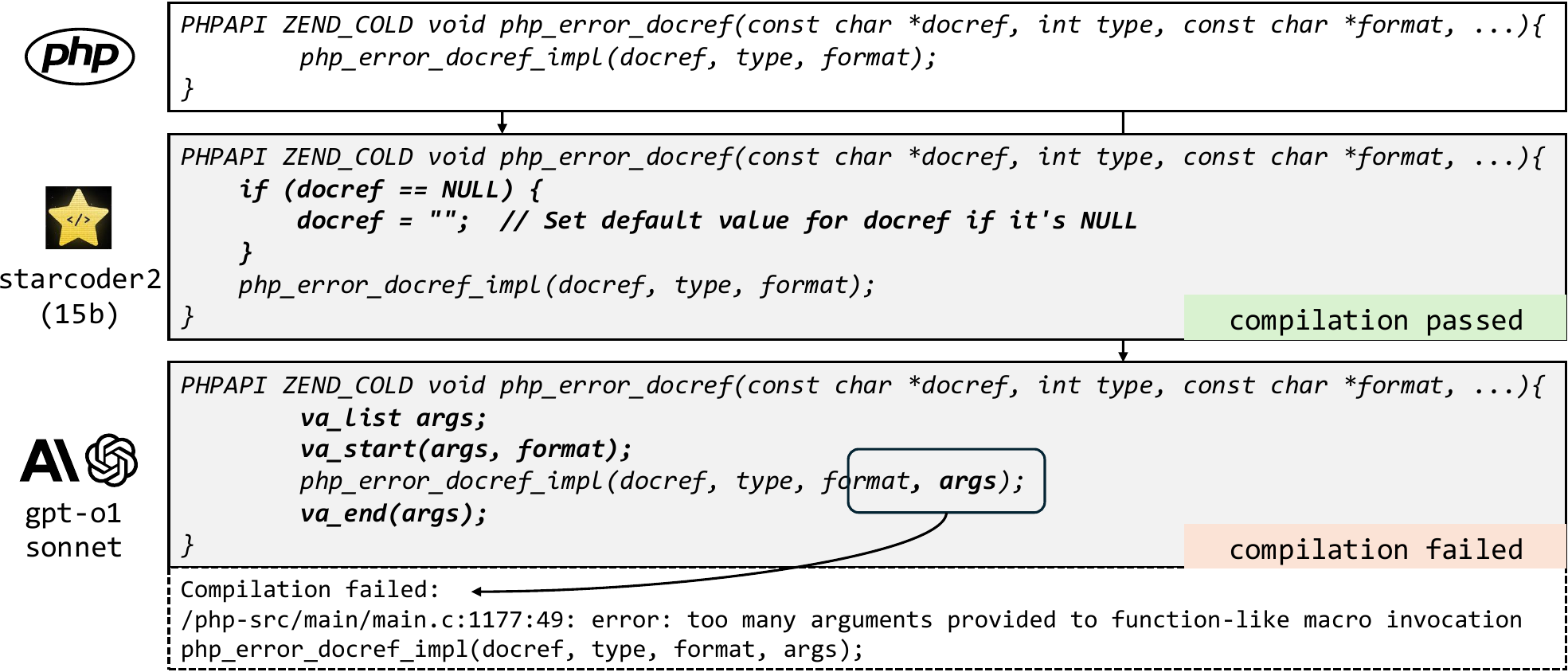}
    \caption{More memorization adds incorrect statements in advanced models}
    \label{fig:variadic}
\end{figure*}

\subsection{Fairness of \name}

We demonstrate that \name can fairly assess LLM coding capabilities even under knowledge contamination. First, we probe each model’s ability to identify OSS function-level snippets and analyze the link between memorization and coding performance. Then, through two case studies, we show that overfitting to these challenging tasks—driven by excessive memorization—can produce incorrect edits in high-end models, whereas lighter models remain robust.

We use a contamination prompt to probe model memorization of OSS source code (details and examples in the Appendix). Our results show that LLMs readily identify a project when explicit identifiers—such as the PHP data structure \textit{zval}—are present. When identifiers are removed, only top-tier models (\texteg GPT-O1, Claude-3.7-Sonnet) occasionally recognize the source, while most models fail on isolated snippets. We observe that higher memorization does not equate to better coding performance: Qwen3-30B-A3B, which recognizes only snippets with obvious keywords, outperforms Gemini-2.5-Flash and Claude-3.5-Haiku (both of which infer sources without PHP-specific identifiers) across all metrics, despite its lower memorization.  

More memorization can sometimes be counterproductive. Figure~\ref{fig:variadic} shows a compilation error produced by some top models (Table~\ref{tab:results}), while others compile the same variadic function in \texttt{main.c} without issue. In their optimized versions, these models injected extra error-handling statements and passed an unintended \texttt{args} parameter, leading to “error: too many arguments provided.” 
% ChatGPT’s explanation\footnote{\url{https://chatgpt.com/share/68076c61-473c-8005-bb5d-fac34c72cf55}} blames the original code rather than the model’s erroneous modification. 
Our analysis of the PHP codebase reveals that the injected statements mirror common error-handling patterns in PHP, explaining why GPT-O1 and Claude-3.7-Sonnet—drawing on their extensive training—introduced them despite breaking this particular function call. Thus, \name mitigates overfitting by leveraging large, complex OSS codebases where memorizing every detail is impractical and can lead to internal inconsistencies.

\begin{figure*}[t]
\setlength{\abovecaptionskip}{0pt}
\setlength{\belowcaptionskip}{0pt}
\setlength{\intextsep}{0pt}
    \centering
    \includegraphics[scale=0.44]{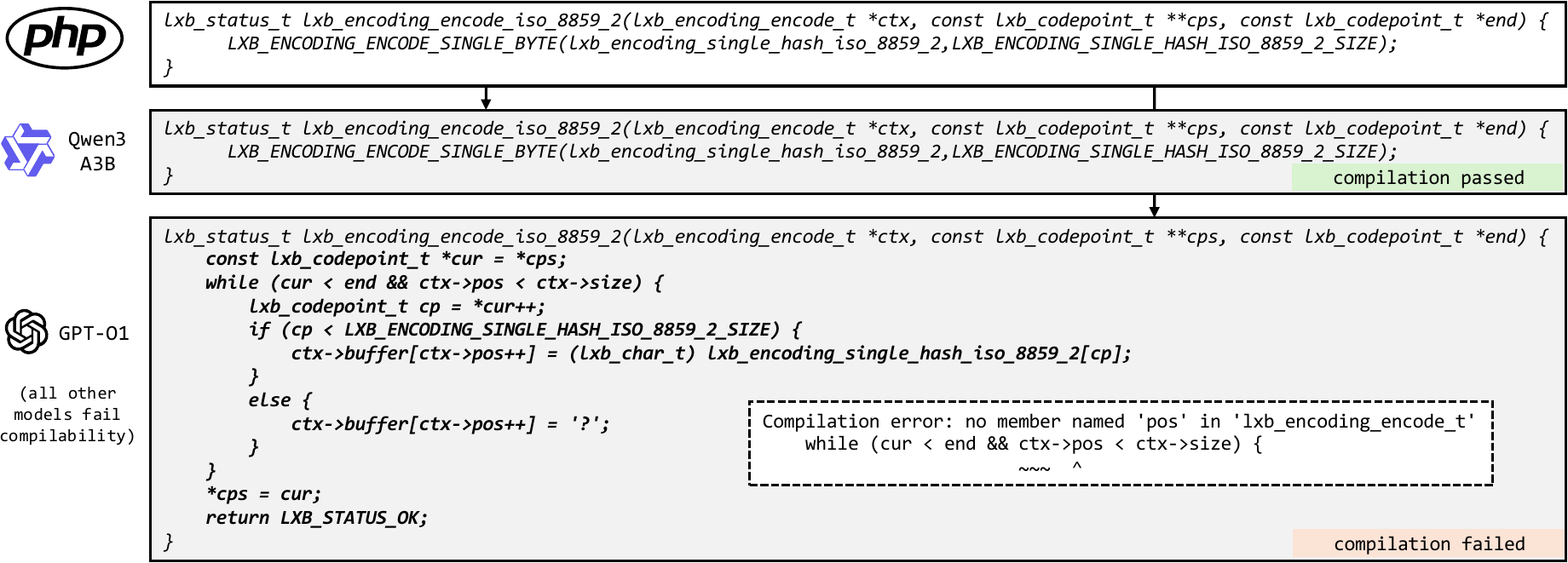}
    \vspace{1mm}
    \caption{Models need caution when generating code for familiar content, such cases may not require optimization, but the presence of elements like \textit{lxb\_encoding} and \textit{iso-8859-2} thills LLMs.}
    % one case that code optimization is not really needed but \textit{iso-8859-2} thrills LLMs}
    \label{fig:iso}
\end{figure*}

Figure~\ref{fig:iso} illustrates one another compilation case in \namephp where only Qwen3-30B-A3B successfully compiles a simple encoding function for \textit{iso-8859-2}. Despite the function’s minimal logic that requires no optimization, other models (\texteg GPT-O1) inject unrelated statements when prompted to improve memory safety and efficiency, causing compilation failures. Qwen3-30B-A3B’s restraint highlights its ability to recognize when no change is needed, avoiding unnecessary edits that break the build. This case highlights that, although language models can memorize extensive patterns, they must exercise restraint in familiar contexts and rethink whether additional modifications are truly necessary according to the user prompt.

We highlight that \name advantage on overfitting mitigation leverages the rapidly evolving nature of open-source software. For example, the PHP project undergoes daily code commits in its actively maintained repository. In such a dynamic environment, overfitting to a static knowledge base becomes impractical, as model memorization can quickly become outdated, potentially leading to incorrect predictions or unexpected failures. Another contributing factor is the use of random sampling in our approach, which generates millions of possible task combinations, making it difficult for models to memorize the exact sampled benchmarks.

% \name greatly mitigates the overfitting problem by leveraging the inherent complexity and difficulty of open-source software benchmarks. To achieve high scores, models must first generate syntactically and semantically correct code. They must then preserve the original functionality and address low-level safety issues—such as overflows. These realistic challenges are difficult even for experienced programmers to solve with confidence.

% In Metric I—Compilability, we observe that higher-performing models do not simply avoid the errors made by lower-performing ones; each model exhibits unique failure cases. 

\subsection{Extended Memory-Safety Evaluation via Fuzzing}

\label{sec:extended_fuzz}

In Section~\ref{sec:approach}, \name measures memory safety by enabling sanitizers at compile time and extracting alerts from compilation and test logs. This approach is efficient, reproducible, and sufficient to distinguish model behaviors. However, compilation and standard tests only cover a limited set of program states, leaving many potential memory bugs undiscovered. Fuzz testing (\texteg randomized input generation) is a well-established method for uncovering deep bugs.

To extend our evaluation, we selected the top-performing closed- and open-source models (GPT-O1 and Qwen3-A3B-30B) along with the PHP baseline and fuzzed the first 500 valid sampled test iterations from Metric II. Using the state-of-the-art PHP fuzzer FlowFusion~\cite{jiang2024fuzzing}, we generated 50,000 mutated inputs per iteration and recorded the unique sanitizer alerts. This extended analysis provides a more rigorous comparison of each model’s memory-safety capabilities.

\begin{table}[!h]
\centering
% \caption{Memory Safety Bug Counts via Extended Fuzzing in \namephp}
\scalebox{0.78}{
\begin{tabular}{lcccccc}\toprule
Models &Param. Size & \#SEGV & \#Spatial & \#Temporal & \#Undefined & Total \\\midrule
\rowcolor{gray!10}
PHP (baseline)           & N/A & 3   & 37  & 1  &  15 & 56 \\
GPT-O1~\cite{gpt-o12025}      & N/A & 101 & 64 & 8 & 163 & 336 \\
Qwen3-A3B~\cite{qwen3-30b-a3b-fp16}   & 30B & 211 & 164 & 11 & 350 & 736 \\
\bottomrule
\end{tabular}
}
\label{tab:fuzz}
\end{table}

The table above presents deduplicated memory safety bug counts via our extended fuzzing. We categorize bugs into four types: \textit{SEGV} (\textie segmentation faults), \textit{Spatial} (\texteg stack or heap overflows), \textit{Temporal} (\texteg use-after-free or double-free), and \textit{Undefined} (other C/C++ undefined behavior, \texteg integer overflows). After 500 fuzzing iterations, even the original PHP code shows exhibits some safety bugs, establishing a realistic baseline. Despite prompts emphasizing memory safety, LLM-generated edits introduce roughly ten times more new violations. Among the evaluated models, GPT-O1 (closed) and Qwen3-A3B-30B (open) produce the fewest new bugs in Table~\ref{tab:results}; all other models generate substantially more. These findings underscore the significant room for improvement in LLMs’ handling of low-level memory safety.  

% \begin{tcolorbox}[colback=gray!5!white,colframe=gray!50!black,title=\textbf{Takeaway}]
% Memory safety is not well considered by LLMs.
% \end{tcolorbox}

\section{Discussion}

\label{sec:discussion}

\textbf{Prompt Flexibility.} We focus exclusively on realistic code-editing prompts for \namephp and \namesql as it is computationally expensive to experiment with various prompts. However, the prompt in \name is flexible and can be extended to more aggressive tasks, such as direct code synthesis or structural transformations (\texteg altering control or data flow).

\textbf{Scoped Scalability.} Although the compilation challenge can be extended to any compiled language, such as C, C++, Rust, Go, Swift, and Java, the sanitizer checks in Metric III are currently limited to memory-safety issues in C/C++ and Rust. While this focus offers a novel dimension for assessing code quality, it constrains the applicability of the sanitizer challenge to other languages and safety domains.

\textbf{Time-Cost Evaluation.} \name typically takes longer evaluation than other benchmarks, driven by the large number of functions in each OSS project and the extensive testing required for robust results. On a typical PC, the evaluation for each model completes in approximately 48 hours for \emph{pass@1}. We consider this time investment reasonable because the evaluation cycle is much shorter than the typical month-level intervals between major LLM updates.

% \textbf{Prompts and code context.} In our current setup, \name’s optimization prompts do not supply additional code context to the LLM since the functions themselves contain the necessary ground-truth information. The use of creative prompts for generating entirely new code within large OSS projects remains unexplored under our framework. 
% \ry{note it would be quite computationally expensive to experiment with many prompts}
% While optimization prompts focus solely on improving and measuring the degradation of existing functions, future work may benefit from incorporating broader code context and more generative prompt designs.

\section{Related Work}

% Recent benchmarks span realistic production and contest-based tasks (BigCodeBench~\cite{zhuo2024bigcodebench}, LiveCodeBench~\cite{jain2024livecodebench}), multilingual and class-level generation (HumanEval‑X~\cite{zheng2023codegeex}, HumanEval‑XL~\cite{peng2024humanevalxl}, ClassEval~\cite{du2023classeval}), practical workflows and library integration (SWE‑Bench~\cite{jimenez2023swebench}, ML‑Bench~\cite{tang2023ml}), domain-specific and instruction-compliant scenarios (DomainCodeBench~\cite{zheng2024well}, CodeIF~\cite{wang2025codeif}), evolution-aware and complexity-controlled generalization (HumanEvo~\cite{zheng2025humanevo}, DynaCode~\cite{hu2025dynacode}, PythonSaga~\cite{yadav2024pythonsaga}, ComplexCodeEval~\cite{feng2024complexcodeeval}), security-focused evaluations (CodeLMSec~\cite{hajipour2024codelmsec}, SecRepoBench~\cite{dilgren2025secrepobench}, CWEval~\cite{peng2025cweval}), and holistic quality and competitive ranking (RACE~\cite{zheng2024beyond}, CodeArena~\cite{baars2019codearena}, CodeElo~\cite{quan2025codeelo}). Further discussion can be found in the Appendix.

Recent benchmarks have increasingly emphasized realistic and dynamic tasks to evaluate the coding capabilities of large language models (LLMs). BigCodeBench~\cite{zhuo2024bigcodebench} assembles realistic programming challenges involving diverse APIs and unit tests to simulate production-level coding, while LiveCodeBench~\cite{jain2024livecodebench} sources tasks from recent coding contests to mitigate data leakage and reduce overfitting risks. HumanEval-X~\cite{zheng2023codegeex} and HumanEval-XL~\cite{peng2024humanevalxl} extend multilingual and cross-lingual code generation capabilities across multiple programming and natural languages, whereas ClassEval~\cite{du2023classeval} evaluates comprehensive class-level Python code generation tasks.

Benchmarks such as SWE-Bench~\cite{jimenez2023swebench,jimenez2023swe} and ML-Bench~\cite{tang2023ml} focus on practical software engineering workflows, targeting real-world GitHub issues, multi-file bug fixes, and integrations with popular machine learning libraries. DomainCodeBench~\cite{zheng2024well} and CodeIF~\cite{wang2025codeif} expand the evaluation scope to domain-specific scenarios and instruction-following tasks, assessing models' adaptability across diverse software ecosystems and adherence to detailed human instructions.

To test generalization and adaptability over software evolution, HumanEvo~\cite{zheng2025humanevo} prompts LLMs with outdated codebase snapshots, requiring forward-compatible code modifications, while DynaCode~\cite{hu2025dynacode} programmatically generates tasks of varying complexity to evaluate true algorithmic reasoning. PythonSaga~\cite{yadav2024pythonsaga} similarly focuses on generalization, examining programming concepts across varying difficulty levels. ComplexCodeEval~\cite{feng2024complexcodeeval} assesses multifaceted tasks from large Java and Python repositories to evaluate broader software development contexts comprehensively.

Security-focused benchmarks, including CodeLMSec~\cite{hajipour2024codelmsec}, SecRepoBench~\cite{dilgren2025secrepobench}, and CWEval~\cite{peng2025cweval}, rigorously evaluate models' secure coding capabilities through vulnerability-aware code generation and static analysis. Holistic quality assessments are represented by RACE~\cite{zheng2024beyond}, which measures code correctness, readability, maintainability, and efficiency, and CodeArena~\cite{baars2019codearena}, which employs adaptive scoring based on collective model performance. Finally, CodeElo~\cite{quan2025codeelo} ranks models on competitive programming tasks using Elo ratings, providing human-comparable skill evaluations.

\section{Conclusion}

We present \name, an automatic benchmark generator that constructs live, large-scale evaluation tasks from real-world open-source software by reusing natural ground truth signals—compilation results, functional test outcomes, and sanitizer checks. Instantiated on the PHP interpreter and SQLite3 engine, \namephp and \namesql can effectively profile the coding capabilities of large language models and reveal insightful patterns, such as intra-family behavior and inconsistencies between model size and performance. Furthermore, \name mitigates overfitting by leveraging the rapidly evolving nature of open-source projects and employing random sampling over dynamic datasets. The framework is scalable across project selection, prompt templates, sampling rates, and metric weights. Future work will extend \name to additional compiled languages like Rust, incorporate richer code context, introduce new metrics such as efficiency profiling, and optimize the fuzzing pipeline to support continued progress in AI-assisted coding tools.

\bibliographystyle{unsrtnat}
\newpage
\bibliography{reference}

\newpage
\section*{Appendix}

\label{appendix:contamination}

\textbf{Contamination Prompt and Example Function.}
The system prompt is \textit{``You are an expert on open-source software''}. The contamination test prompt is formulated as follows: \textit{``Could you recognize the following C function snippet and provide the name of the open-source software it belongs to using the format `Yes. It belongs to ...'? If you are unsure, please respond with `I dont know'. The code snippet is: \{php function\}''}. If the answer contains ``php'', this model is believed to have contaminated memory of the PHP project.

We assess potential memorization by testing LLMs on selected functions from \namephp that omit explicit identifiers such as ``PHP'' or ``Zend''. We demonstrate some function examples as follows:

% We categorize these functions into three difficulty levels based on recognition capability: ``difficult'' (no models correctly identify), ``medium'' (only top models like GPT-O1 and Claude-3.7-Sonnet identify), and ``easy'' (all models recognize successfully).

\lstdefinestyle{compactc}{
  language=C,                                % set language
  basicstyle=\ttfamily\scriptsize,          % smaller monospaced font
  keywordstyle=\color{blue}\bfseries,       % highlight keywords
  commentstyle=\color{gray}\itshape,        % comments in gray italics
  stringstyle=\color{orange},               % strings in orange
  numbers=left,                              % line numbers on left
  numberstyle=\ttfamily\scriptsize,         % line-number font size
  stepnumber=1,                              % show every line number
  numbersep=5pt,                             % distance of numbers from code
  frame=single,                              % thin frame around code
  aboveskip=2pt, belowskip=2pt,              % reduce vertical space
  xleftmargin=5pt, xrightmargin=5pt,         % tighten horizontal margins
  columns=fullflexible,                      % allow tighter character spacing
  keepspaces=true,                           % keep spaces in code
  showstringspaces=false,                    % don't mark spaces in strings
  lineskip=-1pt                              % reduce space between lines
}

\lstset{style=compactc}
\begin{lstlisting}[caption={The PHP function (difficult) that is recognized by no models}]
void KeccakP1600_ExtractBytesInLane(const void *state, unsigned int lanePosition, unsigned char *data, 
unsigned int offset, unsigned int length)
{
    UINT64 lane = ((UINT64*)state)[lanePosition];
#ifdef KeccakP1600_useLaneComplementing
    if ((lanePosition == 1) || (lanePosition == 2) || (lanePosition == 8) || (lanePosition == 12)
    || (lanePosition == 17) || (lanePosition == 20))
        lane = ~lane;
#endif
#if (PLATFORM_BYTE_ORDER == IS_LITTLE_ENDIAN)
    {
        UINT64 lane1[1];
        lane1[0] = lane;
        memcpy(data, (UINT8*)lane1+offset, length);
    }
#else
    unsigned int i;
    lane >>= offset*8;
    for(i=0; i<length; i++) {
        data[i] = lane & 0xFF;
        lane >>= 8;
    }
#endif
}
\end{lstlisting}

\lstset{style=compactc}
\begin{lstlisting}[caption={The PHP function (medium) that is only recognized by GPT-O1 and Claude-3.7-Sonnet}]
static void gc_scan_roots(gc_stack *stack)
{
	uint32_t idx, end;
	gc_root_buffer *current;
	/* Root buffer might be reallocated during gc_scan,
	 * make sure to reload pointers. */
	idx = GC_FIRST_ROOT;
	end = GC_G(first_unused);
	while (idx != end) {
		current = GC_IDX2PTR(idx);
		if (GC_IS_ROOT(current->ref)) {
			if (GC_REF_CHECK_COLOR(current->ref, GC_GREY)) {
				GC_REF_SET_COLOR(current->ref, GC_WHITE);
				gc_scan(current->ref, stack);
			}
		}
		idx++;
	}
	/* Scan extra roots added during gc_scan */
	while (idx != GC_G(first_unused)) {
		current = GC_IDX2PTR(idx);
		if (GC_IS_ROOT(current->ref)) {
			if (GC_REF_CHECK_COLOR(current->ref, GC_GREY)) {
				GC_REF_SET_COLOR(current->ref, GC_WHITE);
				gc_scan(current->ref, stack);
			}
		}
		idx++;
	}
}
\end{lstlisting}

\lstset{style=compactc}
\begin{lstlisting}[caption={The PHP function (easy) that is recognized by all models (due to the use of ``zval'')}]
static void
describe_dict_fn (const char * const lang,
                  const char * const name,
                  const char * const desc,
                  const char * const file,
                  void * ud) /* {{{ */
{
	zval *zdesc = (zval *) ud;
	array_init(zdesc);
	add_assoc_string(zdesc, "lang", (char *)lang);
	add_assoc_string(zdesc, "name", (char *)name);
	add_assoc_string(zdesc, "desc", (char *)desc);
	add_assoc_string(zdesc, "file", (char *)file);
}
\end{lstlisting}

\end{document}